\documentstyle[floats,twocolumn,aps,prb] {revtex}
\begin{document}
\draft
\tighten



\twocolumn[

\begin{center}
\large
{\bf Strain-Dependence of Surface Diffusion: Ag on
Ag\,(111) and Pt\,(111)}
\normalsize

\vspace{0.5cm}

C. Ratsch\cite{Christian}, A. P. Seitsonen, and M. Scheffler

{\it Fritz-Haber-Institut der Max-Planck-Gesellschaft, Faradayweg 4-6,
D-14195 Berlin-Dahlem, Germany}

(Submitted: 2. October 1996)
\end{center}


\vspace{-.9cm}

\begin{center}
\begin{abstract}
\hspace*{1.5cm}\parbox{15.0cm}{

Using density-functional theory with the local-density approximation
and the generalized gradient approximation
we compute the energy barriers for surface diffusion 
for Ag on Pt\,(111), Ag on one monolayer of Ag on Pt\,(111), and Ag on Ag\,(111).
The diffusion barrier for Ag on Ag\,(111) is found to increase linearly
with increasing lattice constant.
We also discuss the reconstruction
that has been found experimentally when two Ag layers are deposited on Pt\,(111).
Our calculations explain why this strain driven reconstruction
occurs only after two Ag layers have been deposited.

\vspace{0.2cm}

PACS number(s): 68.55.Jk, 68.35.Bs, 68.35.Fx

\vspace{0.5cm}

}
\end{abstract}

\end{center}

]

The diffusion of a single adatom on a surface is a fundamental process 
determining
the surface morphology developing in epitaxial growth.
For Ag on Ag\,(111) it has been demonstrated \cite{Rosenfeld}
that the substrate temperature can be lowered periodically for a
short period of time so that diffusion is suppressed, and as a result the density of 
islands is increased
at the beginning of growth of a new layer.
With this trick the growth mode can
be changed from three-dimensional to two-dimensional
(for a discussion of this and other approaches
affecting the growth mode see Ref. \cite{scheffler}).
Growth of one material on a different material is of particular interest for
a number of technological applications. In such a heteroepitaxial system 
the material to be deposited is under the influence 
of epitaxial strain, yet very little is known about the influence of strain 
on the surface diffusion constant. Additionally, strain is 
not only due to lattice mismatch but it is
also present on surfaces of homoepitaxial systems
as a result of the modified bonding configuration.

Experiments show that the diffusion constant follows an 
Arrhenius type behavior 
$D \sim \exp(-E_b/k_BT)$, where $E_b$ is the barrier for surface diffusion and $T$ is 
the substrate temperature.
In a recent scanning tunneling microscopy (STM)
experiment Brune {\it et al.} \cite{Brune95} measured the 
island density $N$ of Ag on Ag\,(111), Ag on Pt\,(111), and Ag on one monolayer (ML) of 
Ag on Pt\,(111) for different temperatures. With the scaling relation \cite{Stoyanov}
$N \sim (D/F)^{-\chi}$, where $F$ is the deposition flux,
the diffusion barrier $E_b$ can be determined as long as the temperature is
low enough \cite{Ratsch} so that the scaling exponent $\chi = 1/3$. Brune {\it et al.}
find that the diffusion barriers are $E_b^{\rm Ag-Pt} = 157\pm10$~meV
for Ag on Pt\,(111), $E_b^{\rm Ag-Ag} =97\pm10$~meV
for Ag on Ag\,(111), and $E_b^{\rm Ag-Ag/Pt} =60\pm10$~meV for Ag on one ML Ag on Pt\,(111).
Thus, the diffusion barrier for Ag on top of a pseudomorphic layer of Ag on Pt\,(111) is substantially lower
than
that for Ag on Ag\,(111).
It has been argued \cite{Goodman} that a metallic monolayer which is 
supported on a dissimilar substrate is electronically perturbed by the substrate, and
that its chemical properties are altered.
The question arises whether the low diffusion
barrier for Ag on one ML Ag
on Pt\,(111) is a result of the compressive strain of $4.2\,\%$ or an electronic effect
due to the Pt underneath the Ag layer.

There are only a few theoretical studies that investigated the effect of lattice mismatch 
on the diffusion barrier. 
In an extensive molecular dynamics study for Si on Si\,(001) that employed a Stillinger-Weber potential 
Roland and Gilmer \cite{Roland} found that the barrier for diffusion along the fast channel
parallel to the dimer rows
is {\it lowered} by approximately $10\,\%$ for both, $3\,\%$ compressive
or $2\,\%$ tensile strain, while diffusion along 
the same direction atop the dimer rows is {\it increased} by about $10\,\%$, so that
a general trend cannot be seen.
Using Lennard-Jones potentials Schroeder and Wolf \cite{Schroeder}
found that the diffusion barrier
increases linearly as the lattice constant increases.
For a metallic system we are only aware of results for Ag on Ag\,(111) where
the authors of  Ref. \cite{Brune95} find in an 
effective medium theory (EMT) calculation that the diffusion barrier increases 
with increasing tensile strain of the surface
and decreases 
with increasing
compressive strain.
Comparison with experiment 
[for Ag on Pt\,(111), Ag on Ag\,(111),  and Ag on one ML Ag on Pt\,(111)]
confirms this trend 
but also shows that the EMT results are off 
with an error between 20\,\% and 100\,\%.
For values of misfit larger than $3\,\%$ the EMT diffusion barrier starts to 
decrease, which is in qualitative disagreement to our results described below.

\vspace{-17.9cm}
\hspace{1.5cm}Physical Review B {\bf 55} (1997), in press
\vspace{17.5cm}

In this work we present first-principles calculations 
of the dependence of the diffusion barrier on 
the lattice constant for Ag on Ag\,(111). This allows us 
to isolate the effect of strain from electronic effects caused by the difference
of materials. 
The dependence of the diffusion barrier on lattice mismatch is 
found to be essentially linear.
The diffusion barriers for the system Ag on Pt\,(111)
are calculated as well.
We find $E_b^{\rm Ag-Pt} = 150$~meV, $E_b^{\rm Ag-Ag} =81$~meV,
and $E_b^{\rm Ag-Ag/Pt} =63$~meV,
which is in good agreement with the
experimental data \cite{Brune95} within the error margins.

We employ density-functional theory (DFT) together with the
local-density approximation (LDA) \cite{LDA} for the exchange-correlation (XC) functional.
The energy barriers are found to be only weakly affected when the 
generalized-gradient approximation (GGA) \cite{GGA} is used instead of the LDA.
In this study the GGA results for the barriers 
are typically higher but not more than $5\,\%$ to $10\,\%$ compared to the LDA results.
Norm-conserving, fully-separable pseudopotentials have been employed that were generated 
according to a scheme proposed by Troullier and Martins.\cite{Pseudo}
The computer code used is described by Bockstedte {\it et al.}  \cite{Michel}.
The calculated bulk lattice 
constants are $a_{\rm Pt}=3.92$~{\AA} for Pt and $a_{\rm Ag}=4.05$~{\AA} for Ag with the 
LDA and $a_{\rm Pt}=4.01$~{\AA} and $a_{\rm Ag}=4.19$~{\AA} with the GGA. 
The LDA results are smaller than the GGA results which is the usual trend.\cite{trend}
For Ag the experimental value is slightly larger (smaller) than the LDA (GGA)
result while for Pt the LDA result is identical to the experimental lattice 
constant. 
In the above theoretical values
the influence of zero-point vibrations
is not included; it would increase
the lattice constant by 
less than $0.2\,\%$.

To simulate the
surface we use the supercell approach.
In the $z$-direction slabs are separated by a vacuum region with a 
thickness that is equivalent to 6 layers, and it has been tested carefully 
that this vacuum region is thick enough.
The adatom is placed on only one side of the slab.
Because of the simple geometry of the (111) surface (cf. Fig. 1)
\begin{figure}[tb]
\unitlength1cm
\begin{center}
   \begin{picture}(10,4.5)
      \includegraphics{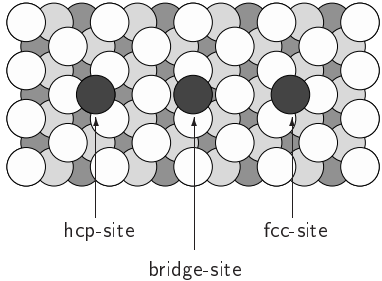}
   \end{picture}
\end{center}
\caption{
A schematic representation of the adsorption sites (fcc and hcp site)
and saddle point (bridge site) on
the (111) surface.
}
\end{figure}

only the fcc and hcp sites need to be considered as adsorption sites with the bridge site as 
the saddle point. It turns out that the fcc site is always slightly favored over 
the hcp site so that all barriers discussed are the differences between the 
total energies of the 
system with the adatom in the
fcc site and the bridge site.

We carried out careful tests varying the slab thickness and surface cell size to
ensure that interactions between neighboring adatoms are negligible.
The tests for Ag on Ag\,(111)
are summarized in Table 1. 
\begin{table}[b]
\caption{Convergence Tests for Ag on Ag\,(111).
GGA is fully self-consistent GGA and GGA-ap is a
posterior GGA as described in the text.
The cell size used was $(2 \times 2)$ except in calculation 7
which was obtained with a $(3 \times 3)$ cell $N_{\bf k}$ is the 
number of ${\bf k}$ points.
}
\begin{tabular}{lllllr}
Calc.  & $E_{\rm cut}$ (Ryd) & $N_{\rm {\bf k}}$ &
$N_{\rm layer}$ & XC & $E_b$ (meV)\\
\hline
1 & 40 & 10 & 3 &  LDA & 82 \\
2 & 40 & 10 & 4 &  LDA & 73 \\
3 & 40 & 10 & 5 &  LDA & 73 \\
4 & 50 & 10 & 4 &  LDA & 81 \\
5 & 60 & 10 & 4 &  LDA & 80 \\
6 & 50 & 8 & 4 & LDA & 81 \\
7 & 40 & 5 & 4 &  LDA & 73 \\
8 & 40 & 10 & 4 &  GGA  & 78 \\
9 & 50 & 10 & 4 &  GGA-ap  & 87 \\
\end{tabular}
\label{table1}
\end{table}
In calculation  1, 2, and 3 we varied the slab thickness $N_{\rm layer}$
and conclude that $N_{\rm layer}=4$ is sufficient.
The electronic wave functions are expanded in plane-waves that are truncated at
a cut-off energy $E_{\rm cut}$. From calculation  2, 4, and 5  
we find that $E_{\rm cut}=40$~Ryd is sufficient if the desired 
accuracy is $\pm 10\,\%$, but for a accuracy of $\pm 2\,\%$ a larger cut-off
$E_{\rm cut}=50$~Ryd is necessary. 
For all the results reported below we choose a slab with a $(2 \times 2)$ cell
as it can be justified from calculation  2 and 7.
We always relaxed the positions of the 
adatom and the atoms of the top layer. A test revealed that the 
results remained unaltered upon relaxation of the second layer.
For the ${\bf k}$-summation we
used 10 ${\bf k}$ points in the irreducible
part of the surface Brillouin zone of the ($2 \times 2$) cell
that were generated according to Ref. \onlinecite{Monkhorst}.
The ${\bf k}$ points included the $\overline{\Gamma}$-point but calculations with different
${\bf k}$ points that did not include the $\overline{\Gamma}$-point
(calculation  6) yielded identical results.
Thus, the error in the Brillouin zone integration is negligible.

Results for the diffusion barrier for Ag on Ag\,(111)
as a function of lattice mismatch are shown in Fig. 2.
\begin{figure}[tb]
\unitlength1cm
\begin{center}
   \begin{picture}(10,4.5)
      \includegraphics{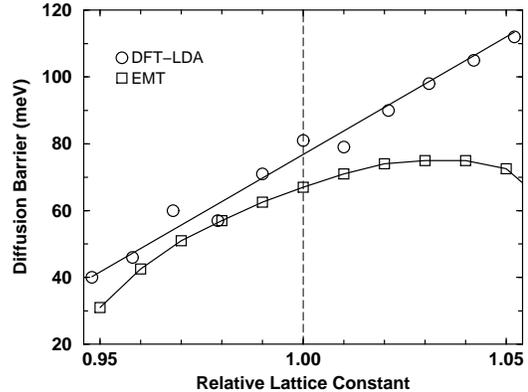}
   \end{picture}
\end{center}
\caption{
The diffusion barrier as a function of the relative lattice constant
$a/a_0$ for Ag on
Ag\,(111).
The lattice constant is normalized to the computed bulk lattice constants
of Ag $a_0 = 4.05$~\AA (DFT) and $a_0 = 4.075$~\AA (EMT).
The EMT results are taken from Ref. 3.
The solid lines are a guide to the eye.}
\end{figure}
For the unstrained system we obtain $E_b^{\rm Ag-Ag} = 81$~meV in good
agreement with the STM results of Ref. \onlinecite{Brune95}.
In the range of the lattice constants studied 
the diffusion barrier $E_b (a/a_0)$ varies linearly within our numerical accuracy \cite{accuracy} with a slope 
of $\sim 0.7$~eV.
This is in contrast to the EMT results of Ref. \onlinecite{Brune95} 
that are also shown in Fig. 2 where the dependence is 
sub-linear and the diffusion barrier decreases for misfits larger than $\sim 3\,\%$.

To further explain the strain dependence of the diffusion barrier
we show in Fig. 3 the strain dependencies of the 
underlying contributions, i.e., the
total energies of an adatom at the fcc hollow and the bridge sites. 
\begin{figure}[tb]
\unitlength1cm
\begin{center}
   \begin{picture}(10,4.5)
      \includegraphics{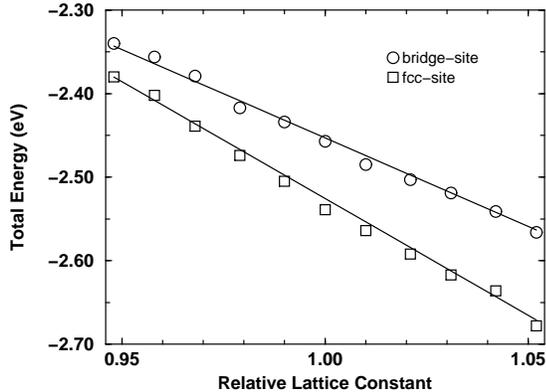}
   \end{picture}
\end{center}
\caption{
The total
energy of an adatom in the fcc site and bridge site
as a function of the relative lattice constant $a/a_0$ for Ag on
Ag\,(111).
The lattice constant is normalized to the computed bulk lattice constants
of Ag $a_0 = 4.05$~\AA \,(DFT) and $a_0 = 4.075$~\AA \,(EMT).
The solid lines are a guide to the eye.
}
\end{figure}
These are computed by subtracting the total energies of a 
clean surface and a free silver atom from the total energy of a surface with
an adsorbed atom.
The total energies  of Fig. 3
show an approximately linear dependence~\cite{accuracy}
within a surprisingly large range of lattice constants.
The change of the
total energy $E_{\rm ad}(a/a_0)$
with the adatom at the threefold coordinated fcc site is 
stronger than that at the twofold coordinated bridge site, and
the slopes are $\sim 2.1$~eV for the bridge site and
$\sim 2.8$~eV for the fcc site. The difference reflects
that the corrugation of the potential energy surface 
gets more pronounced when the surface is expanded 
while under compression it becomes more flat. Obviously,
the value of the
difference equals the value of $0.7$~eV
obtained above from Fig. 2.
We note that  Dobbs {\it et al.} \cite{Dobbs} recently pointed out
that $\Delta E_b = (\sigma_{\rm fcc} -\sigma_{\rm bridge})\epsilon$
where $\epsilon$ is the strain and $\sigma_{\rm fcc}$ ($\sigma_{\rm bridge}$)
is the surface stress with and adatom at the fcc (bridge) site.
The slopes of the curves in Fig. 3 give these
stresses  $\sigma_{\rm fcc}$ and $\sigma_{\rm bridge}$.

The diffusion barrier for Ag on Pt\,(111) has been calculated as $E_b^{\rm Ag-Pt} = 150$~meV
with a $(2 \times 2)$ cell, a slab with 4 layers where the atoms of the top layer and the adatom 
are relaxed, and $E_{\rm cut} = 40$~Ryd. The same 
result has been obtained for {\em i)} a thickness of 3 layers, {\em ii)} the top two layers 
relaxed, and {\em iii)} a cell size of $(3 \times 3)$. We thus conclude that our result is converged.
The value for $E_b^{\rm Ag-Pt}$ is 
in excellent agreement with the STM value of $E_b^{\rm Ag-Pt} = 157 \pm10$~meV.
Previous DFT-LDA results by Feibelman \cite{Feibelman} who
used the Green function theory gave a higher value of $E_b^{\rm Ag-Pt} = 200$~meV.

Our calculations predict that
the diffusion barrier drops dramatically for Ag on Pt\,(111) after one complete layer
of Ag has been deposited. For this diffusion barrier we obtain
$E_b^{\rm Ag-Ag/Pt}=63$~meV in good agreement with the STM value of $E_b^{\rm Ag-Ag/Pt} = 60 \pm10$~meV.
For Ag on one ML of Ag on Pt\,(111) we have chosen $E_{\rm cut}=40$~Ryd and 3 Pt layers. We also 
checked the result with only 2 Pt layers and verified that choosing 3 Pt layers 
is indeed sufficient for this system.
The diffusion barrier for Ag on one ML Ag on Pt\,(111) is almost identical to the 
value of $E_b^{\rm Ag-Ag}=60$~meV we obtained for Ag on Ag\,(111) compressed to the lattice
constant of Pt. This suggests that the small diffusion barrier for Ag on one ML Ag
on Pt\,(111) is mainly due to the effect of strain and not an electronic 
effect because of the Pt underneath.

All the results discussed above have been obtained with the LDA for the 
XC functional. It is an ongoing debate whether the GGA is
really an improvement over the LDA
and we investigated the 
influence of the GGA for all the systems discussed here. For Ag on Ag\,(111) we find that 
the effect of the GGA on the diffusion barriers is only about $5$~meV.
This can be seen in table 1 from calculations 
4 and 8 for a self-consistent GGA according to Ref. \onlinecite{GGA} and from 
calculations 2 and 9 where the energy has been computed 
with the GGA from a self-consistent LDA electron density (GGA-ap) according to Ref. \onlinecite{Martin}.
Similarly the barrier increases by only $5 - 10$~meV for Ag on Pt\,(111) and Ag on one ML Ag
on Pt\,(111). The negligible effect of the GGA is in agreement
with results by Boisvert {\it et al.} for self diffusion of Pt on Pt\,(111) \cite{Guislan} and 
results by Yu and Scheffler for diffusion via the hopping mechanism for Ag on Ag\,(100).\cite{Yu}
The (111) surface is a close packed surface with a very small surface corrugation and
since LDA and GGA results are very close 
it is plausible to assume that for such a system 
LDA and GGA are both good approximations for the exact XC functional.
This situation will presumably be different 
for diffusion events along and especially across steps as it has been found by the 
authors of Ref. \cite{Yu} for Ag on Ag\,(100).

Brune {\it et al.} \cite{Brune95} also measured the island densities of Ag on two ML of
Ag on Pt\,(111) and found that the island density is much larger than it is for Ag
on just one ML of Ag on Pt\,(111). The reason for this increased island density
is not a higher barrier for surface diffusion. The second layer of Ag on Pt\,(111) 
reconstructs in a trigonal network where domains with atoms in the fcc and hcp site alternate.
\cite{Brune94} This reconstruction occurs either during growth with high enough adatom 
mobility or upon annealing and it can be concluded that this trigonal network is 
the equilibrium structure.
The periodicity of these domains is approximately two domain boundaries 
for every 24 atoms. This can be understood
with purely geometrical arguments because the lattice mismatch is 
$\sim 4.2\,\%$ and every domain boundary implies that there is half of an Ag atom 
less, so that the domain network provides an efficient mechanism to relief epitaxial 
strain.
The experiments indicate that these
domain walls act as repulsive walls so that the island density is determined 
by the size of the reconstructed unit cell
and not the barrier for self diffusion on the flat terrace.
It was not clear however why this domain network is formed only after 
2 ML Ag have been deposited and not already upon completion of the first Ag layer.

To answer this question we compared the total energy
of an adatom in the fcc and 
in the hcp site. Calculations were carried out with a 
$(1 \times 1)$ and a $(2 \times 2)$ cell (i.e. coverages $\Theta=1.0$ and $\Theta=0.25$) and slab 
thicknesses of up to 5 layers. We find that the fcc site is energetically more favorable 
than the hcp site in all cases. 
The energy difference 
between those two sites for an Ag adatom on Pt\,(111) is $40$~meV for a $(2 \times 2)$ cell 
($30$~meV for a $(1 \times 1)$ cell), but less than $10$~meV for an Ag adatom on one ML Ag on Pt\,(111).
The reconstruction does not occur in the first layer because the Ag atoms are bound 
much stronger at the fcc sites but does occur in the second layer because the 
total energies for adatoms in the fcc and hcp site are almost indistinguishable.

In conclusion we found that in the range studied here
the diffusion barrier $E_b (a/a_0)$ for Ag on Ag\,(111) increases 
approximately linear with a slope of $\sim 0.7$~eV when the lattice constant increases.
We propose that this result 
might offer a possibility to change the diffusion 
barrier and thus the island density during growth by artificially straining 
the substrate. If this can be done periodically in a controlled manner it
might be an alternative approach to vary the diffusivity {\it in situ}.
Our LDA results
show that careful DFT calculations with the LDA reproduce experimental values
for the surface diffusion barrier
for the system Ag on Pt\,(111) with very high accuracy.
Work to calculate the barriers for diffusion along and across steps 
and across domain boundaries and pre-factors is in progress.

The authors would like to acknowledge helpful discussions with M. Bockstedte, G. Boisvert,
A. Kley, E. Pehlke, P. Ruggerone, and B.D. Yu.


\begin{references}
\bibitem{accuracy}
We estimate the accuracy of our calculation for the total energies
of the adatom at the fcc and bridge site 
and the diffusion barriers to be better than $10$~meV.
For the data at $-2\,\%$ misfit the errors for the total energies
do 
not compensate each other (cf. Fig. 3) so that the value obtained for $E_b$ appears 
to be too low (cf. Fig. 2); however, within the error margins it does not 
violate our assertion that the data can be fitted by a straight line.
\end{references}
\end{document}